\documentclass[10pt]{article}
\usepackage{amsmath,amssymb}

\numberwithin{equation}{section}

\begin{document}

\title{Good and bad tetrads in $f(T)$ gravity}

\author{
N.~Tamanini\footnote{n.tamanini.11@ucl.ac.uk}~ and 
C.~G.~B\"ohmer\footnote{c.boehmer@ucl.ac.uk}\\
Department of Mathematics and Institute of Origins\\
University College London\\ Gower Street, London, WC1E 6BT, UK}

\maketitle

\begin{abstract}
We investigate the importance of choosing good tetrads for the study of the field equations of $f(T)$ gravity. It is well known that this theory is not invariant under local Lorentz transformations, and therefore the choice of tetrad plays a crucial role in such models. Different tetrads will lead to different field equations which in turn have different solutions. We suggest to speak of a good tetrad if it imposes no restrictions on the form of $f(T)$. Employing local rotations, we construct good tetrads in the context of homogeneity and isotropy, and spherical symmetry, where we show how to find Schwarzschild-de Sitter solutions in vacuum. Our principal approach should be applicable to other symmetries as well. 
\end{abstract}

\section{Introduction}

When formulating theories of gravity, the metric tensor is of paramount importance. It contains the information needed to locally measure distances and thus to make theoretical predictions about experimental findings. However, as an alternative dynamical variable, one can use the tetrad $e^a{}_{\mu}$ which is a set of four vectors defining a local frame at every point. The metric and the tetrad are related by
\begin{align}
  g_{\mu\nu} = e^a{}_{\mu} e^b{}_{\nu} \eta_{ab} \,,
\end{align}
where $\eta_{ab}=\mathrm{diag}(+1,-1,-1,-1)$ is the Minkowski metric of the tangent space. The order of the indices in the tetrad is not irrelevant whenever one considers non-diagonal tetrads. In general we have $e^a{}_{\mu}\neq e_{\mu}{}^a$ one being the transposed (matrix) of the other. One immediately notices that $g_{\mu\nu}$ is a scalar under local Lorentz transformations in the tangent space, while $e^a{}_{\mu}$ transforms as
\begin{align}
  e^a{}_{\mu} \mapsto {\Lambda^a}_b \, e^b{}_{\mu} \,,
\end{align}
where ${\Lambda^a}_b$ is a local Lorentz transformation which satisfies
\begin{align}
  \eta_{ac}\,{\Lambda^a}_b\,{\Lambda^c}_d=\eta_{bd} \,.
  \label{004}
\end{align}

Taking the geodesic equation as our starting point, we can realise that gravitational potentials should be encoded in the metric or the tetrad and forces should be contained in their respective derivatives. In the metric approach to gravity one relates the forces to the Christoffel symbols which do not transform as tensors under coordinate transformations. Being guided by the Poisson equation, one arrives rather naturally at the Riemann and Ricci tensors as the basic quantities in the field equations. Following this route one can also construct the Einstein-Hilbert action. When working on a manifold where the connection is not necessarily symmetric, one can easily construct a tensor quantity by considering the skew-symmetric part of the connection. The resulting tensor is the so-called torsion tensor $T^{\sigma}{}_{\mu\nu}$, which can be used to construct a theory known as the Teleparallel Equivalent of General Relativity (TEGR), see~\cite{Unzicker:2005in,tegr}. The Lagrangian of this formulation differs from general relativity only by a surface term.

Gravitational theories built from the metric and quantities derived from it will always be Lorentz scalars and such theories will, by definition, be invariant under local Lorentz transformations. On the other hand, when building theories with torsion, those would not in general be invariant under local Lorentz transformations since
\begin{align}
  T^{\sigma}{}_{\mu\nu} = \Gamma^{\sigma}{}_{\nu\mu} - \Gamma^{\sigma}{}_{\mu\nu} =
  e_i{}^{\sigma} (\partial_\mu e^i{}_{\nu}-\partial_\nu e^i{}_{\mu}) \,, 
\end{align} 
and therefore
\begin{align}
  T^{\sigma}{}_{\mu\nu} \mapsto T^{\sigma}{}_{\mu\nu} + \Lambda_a{}^b e_b{}^\sigma \left( e^c{}_\nu \partial_\mu\Lambda^a{}_c -e^c{}_\mu \partial_\nu\Lambda^a{}_c \right) \,,
\end{align}
under local Lorentz transformations in the tangent space.
When considering an action based on quadratic combinations of $T^{\sigma}{}_{\mu\nu}$, local Lorentz invariance may be achieved by fine tuning the model. In general, theories like $f(T)$ gravity are not invariant under local Lorentz transformations~\cite{Li:2010cg,Sotiriou:2010mv}, unless of course $f(T) = c_1 + c_2 T$, where $c_1$ and $c_2$ are constants. Conversely, this means that a non-invariant theory will be sensitive to the choice of the tetrad and different tetrads might give rise to different solutions. As such, the choice of the tetrad is a crucial and rather subtle point when studying such theories.  

In other words, although local Lorentz transformations do not change the metric, they do change the $f(T)$ field equations. This happens precisely because the $f(T)$ Lagrangian is not invariant under such kind of transformations. Every different choice of a tetrad giving back the same metric will then represent a different physical theory, describing different modifications of TEGR. However, if we require that for physically viable models TEGR is recovered in some limit, all these possible theories must coincide in such a limit since the TEGR action is invariant under local Lorentz transformations.

In the present paper we study the issue of choosing suitable tetrads within the framework of $f(T)$ gravity. We state that good tetrads do not imply restrictions on the functional form of $f(T)$. The corresponding modified gravitational theories will share solutions with GR and will in turn be considered as better viable modifications of GR.

In Sec.~\ref{Sec:SpherSymm} we analyse spherically symmetric spacetimes. We prove Birkhoff's theorem in full generality and show how, employing local rotations, we can find a good tetrad leading to Schwarzschild-de Sitter solutions. In Sec.~\ref{Sec:Cosmo} we use the same method to build good tetrads for FLRW cosmology in spherical coordinates, first in the flat case and then in the curved ones. Finally, we recap the major results and draw conclusions in Sec.~\ref{Sec:Concl}

\section{Rotated tetrads in spherical symmetry}
\label{Sec:SpherSymm}

Spherically symmetric spacetimes within $f(T)$ gravity have recently received substantial attention~\cite{Wang:2011xf,Deliduman:2011ga, Bohmer:2011si,HamaniDaouda:2011iy,Daouda:2011rt,Daouda:2012nj, Nashed:2011fz,Meng:2011ne,Dong:2012en, Boehmer:2011gw,Ferraro:2011ks,Iorio:2012cm}. Several vacuum and non-vacuum solutions have been built but the Schwarzschild solution has only been found in isotropic coordinates with the aid of a boosted tetrad~\cite{Ferraro:2011ks}. Birkhoff's theorem has been proved using a diagonal tetrad~\cite{Meng:2011ne} which constrains the torsion scalar to be constant and does not admit the Schwarzschild solution~\cite{Boehmer:2011gw}. In what follows we show how employing non-diagonal (rotated) tetrads permits to recover Schwarzschild-de Sitter (SdS) solutions in vacuum and to prove Birkhoff's theorem in full generality, in strictly analogy with \cite{Dong:2012en}.

Consider the general (non-static) spherically symmetric metric
\begin{align}
  ds^2=e^{A(t,r)}dt^2-e^{B(t,r)}dr^2-r^2d\Omega^2 \,,
  \label{001}
\end{align}
where $d\Omega^2=d\theta^2+\sin^2\theta d\phi^2$. The simplest possible tetrad giving this metric is the diagonal one
 \begin{align}
   {e_\mu}^a|_{\mbox{diag}} =
   \begin{pmatrix}
     e^{A(t,r)/2} & 0 & 0 & 0 \\
     0 & e^{B(t,r)/2} & 0 & 0 \\
     0 & 0 & r & 0 \\
     0 & 0 & 0 & r\sin\theta
   \end{pmatrix} \,.
   \label{005}
 \end{align}
Using this tetrad it has been shown that every spacetime described by metric~(\ref{001}) has to be static in $f(T)$ gravity~\cite{Meng:2011ne}. However, the Schwarzschild solution is not a solution of the $f(T)$ field equations derived from tetrad~(\ref{005}) \cite{Boehmer:2011gw} implying that even if Birkhoff's theorem holds, vacuum solutions of the theory do not reduce to GR vacuum solutions as one would expect.

As mentioned in the previous section, we can always change tetrad~(\ref{005}) without affecting metric~(\ref{001}) by a local Lorentz transformation in the tangent space
\begin{align}
  {e_\mu}^a|_{\mbox{diag}}\mapsto {e_\mu}^a=\Lambda_b{}^a\,{e_\mu}^b|_{\mbox{diag}} \,.
\label{002}
\end{align}
In the forthcoming analysis we reduce the local Lorentz transformation matrix to a general 3-dimensional rotation $\mathcal{R}$ parametrised by its three Euler angles $\varphi$, $\vartheta$, $\psi$ such that we can write
\begin{align}
  {\Lambda^a}_b=
  \begin{pmatrix}
    1 & 0 \\
    0 & \mathcal{R}(\varphi,\vartheta,\psi) \\
  \end{pmatrix}\,,
  \label{003}
\end{align}
where
\begin{align}
  \mathcal{R}(\varphi,\vartheta,\psi) = R_z(\psi) R_y(\vartheta) R_x(\varphi)\,,
  \label{003a}
\end{align}
where the $R_x$, $R_y$, $R_z$ are the rotation matrices about the Cartesian coordinate axis with angles $\varphi$, $\vartheta$, $\psi$, respectively. These well-known matrices are given by
\begin{align}
  R_x(\phi) &= 
  \begin{pmatrix}
    1 & 0 & 0 \\
    0 & \cos\varphi & -\sin\varphi \\
    0 & \sin\varphi & \cos\varphi
    \end{pmatrix}\,,\quad
  R_y(\vartheta) = 
  \begin{pmatrix}
    \cos\vartheta & 0 & -\sin\vartheta \\
    0 & 1 & 0 \\
    \sin\vartheta & 0 & \cos\vartheta
  \end{pmatrix}\,,
  \nonumber \\
  R_z(\psi) &=
  \begin{pmatrix}
    \cos\psi & -\sin\psi & 0 \\
    \sin\psi & \cos\psi & 0 \\
    0 & 0 & 1
  \end{pmatrix}\,.
\end{align}

In general, even if $\varphi$, $\vartheta$, $\psi$, are taken to be arbitrary functions of the spherical coordinates $t$, $r$, $\theta$, $\phi$, the transformed tetrad~(\ref{002}) returns metric~(\ref{001}). Of course, this happens because the rotation matrix~(\ref{003}) is a local Lorentz transformation satisfying condition~(\ref{004}). Note that the spherical symmetry of the spacetime is not affected by transformation~(\ref{003}) since it operates within the tangent space. This process is similar to the one used in~\cite{Ferraro:2011ks} where the diagonal tetrad (in isotropic coordinates) is boosted, instead of rotated, by a local Lorentz transformation.

For our purposes, we will consider the following values for the three Euler angles:
\begin{align}
  \varphi = \gamma(r) \,,\quad
  \vartheta = \theta-\pi/2\,, \quad 
  \psi = \phi\,.
\end{align}
where $\gamma$ is taken to be a general function of $r$. With these values the local rotation~(\ref{003}) becomes
\begin{align}
  {\Lambda^a}_b=
  \begin{pmatrix}
    1 & 0 \\
    0 & \mathcal{R}(\gamma(r),\theta-\pi/2,\phi)
  \end{pmatrix}\,,
\end{align}
where we have
\begin{align}
  \mathcal{R}(\gamma(r),\theta-\pi/2,\phi) = R_z(\phi) R_y(\theta-\pi/2) R_x(\gamma(r))\,,
\end{align}
and the rotated tetrad reads
\begin{multline}
{e_\mu}^a=
\left(
\begin{array}{cc}
 e^{A/2} & 0 \\
 0 & e^{B/2} \sin \theta \cos \phi  \\
 0 & -r \left(\cos \theta \cos \phi \sin \gamma+\sin \phi \cos\gamma \right) \\ 
 0 & r \sin \theta \left(\sin \phi \sin\gamma -\cos \theta \cos \phi \cos\gamma \right)
\end{array}\right.\\
\left.
\begin{array}{cc}
 0 & 0 \\
 e^{B/2} \sin \theta \sin \phi  & e^{B/2} \cos \theta  \\
r \left(\cos \phi \cos\gamma -\cos \theta \sin \phi \sin\gamma \right) & r
   \sin \theta \sin\gamma  \\
 -r \sin \theta \left(\cos \theta \sin \phi \cos\gamma +\cos
   \phi \sin\gamma \right) & r \sin ^2\theta \cos\gamma 
\end{array}
\right) \,.
\label{006}
\end{multline}
We notice that setting $\gamma(r) = -\pi/2$ this tetrad reduces to the rotated tetrad considered in~\cite{Boehmer:2011gw} and \cite{Dong:2012en}. This particular choice will turn out useful later when we will prove the staticity of metric~(\ref{001}) without having any constraint on the torsion scalar $T$ (see Sec.~\ref{appA}).

The torsion scalar obtained from tetrad~(\ref{006}) is
\begin{multline}
  T(r) = \frac{2\, e^{-B}}{r^2} 
  \Bigl[ 
    1 + e^B + 2\,e^{B/2} \sin\gamma + 
    2\, e^{B/2}\, r\,  \gamma' \cos\gamma \\+
    r\, A' \left(1+e^{B/2}\sin\gamma\right) 
  \Bigr]\,,
  \label{008}
\end{multline}
where a prime denotes derivative with respect to $r$. The $f(T)$ field equations result in 6 independent relations
\begin{multline}
  4\pi\rho = \frac{f}{4} -\frac{f_T\,e^{-B}}{4r^2}\left( 2-2\,e^B+r^2e^B T-2r\,B' \right) \\
  -\frac{f_{TT}\,T'e^{-B}}{r} \left(1+e^{B/2}\sin\gamma\right) \,,
\label{025}
\end{multline}
\begin{align}
  4\pi p &= -\frac{f}{4} +\frac{f_T\,e^{-B}}{4r^2} \left( 2-2\,e^B+r^2e^B T-2r\,A' \right) \,,
\label{026}
\end{align}
\begin{align}
  f_{TT}\,T'\cos\gamma &= 0 \,, \label{007}\\
  f_T\,\dot B &= 0 \,, \label{009}
\end{align}
\begin{multline}
  \dot{B} \left[ e^B r^2 f_T + 2\,f_{TT}\left(1+e^{B/2}\sin\gamma\right) \left(2-2\,e^B+r^2e^B T+ 2r\, A'\right) \right]\\
  - 4r\,f_{TT}\,\dot{A}'\left( 1+e^{B/2}\sin\gamma\right)^2 =0 \,,
\label{024}
\end{multline}
\begin{multline}
  f_{TT}\Bigl[ -4\,e^A r\, T' -\dot{B}^2\left( 2-2\,e^B+r^2e^B T \right) -2r\,A' \left(e^Ar\,T'+\dot B^2\right) \\
  +4r \dot B\dot A' \left(1+e^{B/2}\sin\gamma\right) \Bigr]
  +f_T \Big[ 4\,e^A-4e^Ae^B-e^Ar^2A'^2+ 2\,e^Ar\,B'\\
  +e^Ar\,A' \left( 2+r\,B' \right) -2r^2e^A A''
  - e^Br^2 \dot A\dot B+e^Br^2 \dot B^2 +2e^Br^2\ddot B \Big] =0\,,
\label{027}
\end{multline}
where $f_T$ and $f_{TT}$ are the first and second derivatives of $f(T)$ and overdots denote differentiation with respect to $t$.

\subsection{Birkhoff's theorem}

In this section we will use the field equations derived above to prove Birkhoff's theorem in $f(T)$ gravity.

\subsubsection{The case $T'=0$}
\label{Sec:T'=0}

Let us start looking at~(\ref{007}). If $\cos\gamma =0$ this is identically satisfied and we gain no constraints on the $r$-dependence of $T$. This case is of particular interest and is analysed in Sec.~\ref{appA}. For the moment we require $\cos\gamma\neq 0$. Moreover since we already know that Birkhoff's theorem holds in GR, or equivalently in TEGR~\cite{Bakler:1980fe}, we can exclude the case $f_{TT}=0$ which would lead back to TEGR. With these premises, condition~(\ref{007}) implies
\begin{align}
  T' = 0 \,,
  \label{010}
\end{align}
meaning that the torsion scalar~(\ref{008}) has to be a constant $T=T_0$.

At this point we can look at~(\ref{009}). Because of~(\ref{010}), $f_T=f_{T_0}$ is now a constant and nothing prevents it to vanish. For example in the model $f(T) = T-T^2/(2T_0)$ we can have $T'=T_0$, $f_{TT}\neq 0$ and $f_{T_0}=0$ satisfying both~(\ref{007}) and~(\ref{009}). Note that the condition $f_{TT}=0$ can be achieved only within TEGR since for a general non-constant $T$ this represents a constraint on the $f(T)$ model to choose. We have to exclude these peculiar models from our analysis. This is not a significant restriction since if $f_{T_0}=0$ the field equations are always satisfied leading to constant $\rho$ and $p$ and a dark energy equation of state: $\rho=-p=f_0/4$, with $f_0=f(T_0)$. It is then unlikely that interesting physical insights can arise from such models.

We then assume $f_{T_0}\neq 0$ and~(\ref{009}) directly yields
\begin{align}
\dot B=0 \,.
\label{011}
\end{align}
Taking into account conditions~(\ref{010}) and~(\ref{011}), the field equations reduce to
\begin{align}
  & 4\pi\rho = \frac{f_0}{4} -\frac{f_{T_0}\,e^{-B}}{4r^2}\left( 2-2\,e^B+r^2e^B T_0-2r\,B' \right) \label{013} \,,\\
  & 4\pi p= -\frac{f_0}{4} +\frac{f_{T_0}\,e^{-B}}{4r^2} \left( 2-2\,e^B+r^2e^B T_0-2r\,A' \right) \label{014} \,,\\
  & \dot A' \left(1+e^{B/2}\sin\gamma\right)^2 =0 \label{012} \,,\\
  & 4-4\,e^B -r^2A'^2 +2r\,B'+r\,A'\left(2+r\,B'\right) -2r^2A'' =0 \,. \label{015}
\end{align}
Consider equation~(\ref{012}), this is satisfied either if $\dot A'=0$ or $1+e^{B/2}\sin\gamma =0$. The second case will be analysed later in Sec.~\ref{App1}, for the moment let assume $1+e^{B/2}\sin\gamma \neq 0$. The constraint $\dot A'=0$ implies that $A(t,r)$ can only be of the type
\begin{align}
  A(t,r) = \psi(r) +\xi(t) \,,
\end{align}
where $\psi$ and $\xi$ are general functions of $r$ and $t$ respectively. Now note that in the field equations~(\ref{013}),~(\ref{014}) and~(\ref{015}) the metric function $A$ appears only through its derivatives with respect to $r$, namely $A'$ and $A''$. This means that this system of equations is invariant under the transformation $A(t,r)\rightarrow A(t,r) +\chi(t)$, with $\chi$ arbitrary function of $t$. We can then always choose $\chi(t)=-\xi(t)$ without affecting the field equations and thus the physical properties of the system. This in turn leads to
\begin{align}
  A(t,r)=\psi(r)\equiv A(r) \,.
\end{align}
Thus both the metric functions are independent of time and the spacetime becomes static:
\begin{align}
  ds^2=e^{A(r)}dt^2-e^{B(r)}dr^2-r^2d\Omega^2 \,.
\label{022}
\end{align}

Alternatively we could get rid of the time dependence of $A$ through a rescaling of the time coordinate. In fact taking $dt\rightarrow e^{\xi(t)/2}dt$ would have given again the static metric~(\ref{022}). With this last approach the physical meaning of this operation is probably more evident. The result we just derived for $T'=0$ is exactly the one found in~\cite{Meng:2011ne} with a diagonal tetrad. We will see in Sec.~\ref{Sec:Scw-deSitt-Sol} that with a non-diagonal (rotated) tetrad this analysis can be pushed further to prove that the most general spherically symmetric solution has to be of the SdS kind. Finally we note that in this case Birkhoff's theorem holds even in presence of matter since we did not assume the vacuum condition nowhere. As we are going to see this is not true in the other cases.

\subsubsection{The case $e^{B/2}\,\sin\gamma+1=0$}
\label{App1}

As said above equation~(\ref{012}) is satisfied either if $\dot A'=0$ or $1+e^{B/2}\sin\gamma =0$. We now consider the second case where $\sin\gamma(r)$ takes the particular form
\begin{align}
  \sin\gamma(r)=-e^{-B/2} \,.
  \label{034}
\end{align}
In the field equations~(\ref{013}),~(\ref{014}) and~(\ref{015}) we still have the time dependency of $A(t,r)$, while $B$ is a function of only $r$ because of~(\ref{011}). The torsion scalar~(\ref{008}) reduces to
\begin{align}
  T_0= \frac{2\,e^{-B}}{r^2}\left( -1+e^B+r\,B' \right) \,,
\end{align}
which can be easily integrated to give
\begin{align}
  e^{-B(r)}=1-\frac{2\,M}{r}-\frac{T_0}{6}r^2 \,,
\end{align}
with $M$ being an integration constant. The field equations then reduce to
\begin{align}
  4 \pi \rho &=\frac{f_0}{4}\,,\label{020}\\
  4 \pi p &=-\frac{f_0}{4}-\frac{f_{T_0}}{12 r^3} \left[r A' \left(12 M+r^3 T_0-6 r\right)+12 M-2 r^3 T_0\right] \,,
\end{align}
\begin{multline}
  2 \left[r^2 A'' \left(12 M+r^3 T_0-6 r\right)-36 M\right] \\
  +r^2 \left(A'\right)^2 \left(12 M+r^3 T_0-6 r\right)+12 r A' (r-3 M) =0 \,. \label{019}
\end{multline}
We can immediately notice that in this case the energy density has to be constant. This automatically excludes all the solutions describing physical models where $\rho \neq $ const. However, though this solution seems to be strongly limited, we can still proceed with our analysis.

Equation~(\ref{019}) is satisfied if $A(t,r)=-B(r)+\xi(t)$, where $\xi$ is again a general function of $t$. Note that this is just one solution to equation~(\ref{019}) and in general there could be other possible solutions. This means that in this particular case Birkhoff's theorem cannot be proved in full generality (unless one shows that this is the unique solution of equation~(\ref{019})). 

However, if we restrict our analysis to the vacuum case $\rho=p=0$, adding equations~(\ref{013}) and~(\ref{014}) leads directly to $A'+B'=0$, and thus to the relation $A(t,r)=-B(r)+\xi(t)$. Note also that in vacuum we must have $f_0=0$ because of~(\ref{020}). Then, noting again that $A(t,r)$ enters the field equations~(\ref{020})--(\ref{019}) only through its derivatives with respect to $r$, the field equations are invariant under the transformation $A(t,r)\rightarrow A(t,r)+\chi(t)$, for any arbitrary function $\chi$ of $t$. Choosing $\chi(t)=-\xi(t)$ we have
\begin{align}
  A(t,r)=A(r)=-B(r)=\log\left(1-\frac{2\,M}{r}-\frac{T_0}{6}r^2\right) \,,
\end{align}
and the spacetime is again static with the metric
\begin{align}
  ds^2=e^{A(r)}dt^2-e^{-A(r)}dr^2-r^2d\Omega^2 \,,
\end{align}
where
\begin{align}
  e^{A(r)}=1-\frac{2\,M}{r}-\frac{T_0}{6}r^2 \,.
\label{021}
\end{align}
This is a SdS spacetime, where the cosmological constant coincides with the value of $T_0/2$ and does not depend on the $f(T)$ model one chooses. Again an alternative way to get rid of the time dependency of $A$ is through a rescaling of time exactly as it happens in the general case.

In conclusion, even in the case $e^{B/2}\,\sin\gamma+1=0$, Birkhoff's theorem can be enforced in vacuum and leads directly to a Schwarzschild-de Sitter solution. Unfortunately, though the static case is a solution also of the general field equations with matter, Birkhoff's theorem cannot be fully proved in the presence of matter since other non-static solutions could be found. However, the fact that the energy density is constrained to be constant, raises many doubts on the physical viability of this particular case.

\subsubsection{The case $\cos\gamma=0$}
\label{appA}

We now go back to~(\ref{007}) and analyse the case $\cos\gamma=0$.
In order to satisfy $\cos\gamma=0$ we must have $\gamma=\pm\pi/2+2n\pi$ with $n$ any integer number. However we will consider $n=0$ in the following since no physical differences arise when $n\neq 0$. In this case~(\ref{007}) is satisfied and tetrad~(\ref{006}) reduces to
\begin{multline}
  {e_\mu}^a=
  \left(
  \begin{array}{cc}
    e^{\frac{1}{2} A(t,r)} & 0 \\
    0 & e^{\frac{1}{2} B(t,r)} \sin \theta \cos \phi  \\
    0 & \mp r \cos \theta \cos \phi \\ 
    0 & \pm r \sin \theta \sin \phi
  \end{array}\right.\\
  \left.
  \begin{array}{cc}
    0 & 0 \\
    e^{\frac{1}{2} B(t,r)} \sin \theta \sin \phi  & e^{\frac{1}{2} B(t,r)} \cos \theta  \\
    \mp r \cos \theta \sin \phi & \pm r
    \sin \theta \\
    \mp r \sin \theta \cos
    \phi & r \sin ^2\theta \cos (\gamma (r))
  \end{array}
  \right) \,,
  \label{023}
\end{multline}
where here, and in the following equations, the upper sign has to be taken for $\gamma=\pi/2$ and the lower sign for $\gamma=-\pi/2$. We stress that when $\gamma=-\pi/2$ tetrad~(\ref{023}) becomes the off-diagonal tetrad considered in~\cite{Dong:2012en,Boehmer:2011gw}.
The torsion scalar reduces to
\begin{align}
  T(t,r) = \frac{2 e^{-b} }{r^2}\left(e^{b/2}\pm 1\right) \left(e^{b/2}\pm r a_r\pm 1\right)\,,
  \label{028}
\end{align}
and field equation~(\ref{009}) directly yields
\begin{align}
  \dot B=0 \,,
\end{align}
since we exclude the unphysical $f_T=0$ model. Looking then at~(\ref{024}) we find
\begin{align}
  f_{TT}\,\dot{A}'\left( 1\pm e^{B/2}\right)^2 = 0 \,,
  \label{033}
\end{align}
which gives rise to three possibilities.

First, equation~(\ref{033}) is satisfied if $f_{TT}\equiv 0$ so that $f(T)= c_1T+c_2$ for $c_1$ and $c_2$ constants. This condition leads to TEGR where Birkhoff's theorem is already valid. Hence to avoid such limitations we will assume $f_{TT}\neq 0$ and look at the remaining two cases, namely $\dot A'=0$ and, only in the $\gamma=-\pi/2$ eventuality, $B=0$.

If $\dot A'=0$ the function $A$ must assume the following form: $A(t,r)=\phi(r)+\xi(t)$, where $\phi$ and $\xi$ are general functions only of $r$ and $t$ respectively. Again if we now try to substitute $A$ into the field equations~(\ref{025}),~(\ref{026}) and~(\ref{027}), we notice that the function $\xi(t)$ completely disappears since $A$ enters these equations only through its derivative with respect to $r$. This suggests that the field equations are invariant under a (gauge) transformation of the kind $A(t,r)\rightarrow A(t,r)+\psi(t)$, with $\psi$ a general function of $t$. In particular we can always make such a transformation with $\psi(t)=-\xi(t)$, all without modifying the field equations. In this manner we remain with $A(t,r)=\phi(r)\equiv A(r)$ meaning that Birkhoff's theorem holds in this particular case with a general metric of the kind
\begin{align}
  ds^2=e^{A(r)}dt^2-e^{B(r)}dr^2-r^2d\Omega ^2\,.
\end{align}
We have just proved that tetrad~(\ref{023}) permit to validate Birkhoff's theorem without imposing any constraint on the torsion scalar $T$. This result generalises both the ones found in Sec.~\ref{Sec:T'=0} and~\cite{Meng:2011ne} where $T$ was constrained to be constant.
Moreover it matches the result found in \cite{Dong:2012en} where Birkhoff's theorem was proven in the $\gamma=-\pi/2$ case.

Finally, in the eventuality where $\gamma=-\pi/2$ and $B=0$, we can look back at the torsion scalar~(\ref{028}) and notice that this gives $T=0$. The remaining field equations reduce to
\begin{align}
  4\pi\rho &= \frac{f(0)}{4} \label{029} \,,\\
  4\pi p &= -\frac{f(0)}{4}-\frac{f_T(0)}{2r}A' \label{032}\,,\\
  0 &= 2A'-r\,A'^2-2r\,A'' \label{030} \,.
\end{align}
Equation~(\ref{029}) constrains the energy density to be constant. This is analogous to what we saw in Sec.~\ref{App1}, meaning that this case can be of interest only for particular physical systems where $\rho=$ const.
The field equations~(\ref{029})--(\ref{030}) have been solved in~\cite{Boehmer:2011gw} for a static spacetime ($A=A(r)$). We have now also time dependency and the solution found in~\cite{Boehmer:2011gw} can be generalised as
\begin{align}
  A(t,r)=2 \log \left[r^2- \xi (t)\right]+2\,\psi (t) \,,
\label{031}
\end{align}
where $\xi$ and $\psi$ are arbitrary function of $t$. We are not stating that this is the general solution of~(\ref{031}), but only one solution which we use to explain what follows. Note that even if we can get rid of $\psi$ with a time rescaling as $t\mapsto e^\psi t$, we cannot avoid the time dependency of $A$ through $\xi$ in~(\ref{031}). Thus in this case Birkhoff's theorem is not satisfied in general. However in vacuum we can add~(\ref{029}) and~(\ref{032}) which immediately gives $A'=0$, meaning that $A(t,r)=A(t)$. Equation~(\ref{030}) is thus identically satisfied and we must have $f(0)=0$. We can now rescaling the time coordinate as $t\mapsto e^A t$ in the metric (which corresponds to choose the gauge such that $A=0$). This leads to Minkowski spacetime $g_{\mu\nu}=\eta_{\mu\nu}$ and thus, being it static, to the validity of Birkhoff's theorem.

In conclusion, analysing all the cases arising from field equations~(\ref{025})--(\ref{027}), we found that Birkhoff's theorem is generally valid in $f(T)$ gravity. In the case where $\cos\gamma=0$ this has been proved without imposing any constraint neither on the Lagrangian function $f(T)$ nor on the torsion scalar $T$. This is in agreement with \cite{Dong:2012en} and generalised the result found in~\cite{Meng:2011ne} where Birkhoff's theorem was proved employing a diagonal tetrad which leads to the $T'=0$ constraint. Furthermore, using the rotated tetrad~(\ref{006}), if $T'=0$ not only is possible to prove Birkhoff's theorem even in the presence of matter, but it is also possible to show that the most general vacuum solution has to be of the SdS kind. This is what we present in the following section.

\subsection{Schwarzschild-de Sitter solutions}
\label{Sec:Scw-deSitt-Sol}

In this section we consider again the $T'=0$ case for which we proved that metric~(\ref{001}) has to be static and reduce to~(\ref{022}). We will show that in vacuum the most general solution to the field equations has to be a Schwarzschild-de Sitter (SdS) solution. We have already seen that in the case exposed in Sec.~\ref{App1} the vacuum solution is of this type. In what follows we prove that even when the (unphysical) condition~(\ref{034}) is not satisfied this statement is true.

Using metric~(\ref{022}) the field equations are given by equations~(\ref{013}),~(\ref{014}) and~(\ref{015}) where now both $A$ and $B$ are functions of $r$ only. Note that in these equations the function $\gamma(r)$ completely drops out exactly as it happens in the field equations given by the boosted tetrad considered in~\cite{Ferraro:2011ks}. This function plays the role of an auxiliary function whose value does not affect the physical properties of the system. Once the metric functions $A$ and $B$ are known, $\gamma(r)$ is always determined by equation~(\ref{008}) with $T(r)=T_0$ parametrising all the possible solutions. Thus we only have to find the expressions for $A(r)$ and $B(r)$ since the function $\gamma(r)$ will follow from~(\ref{008}).

Now consider the sum of~(\ref{013}) and~(\ref{014}) which yields (note that this can also be done with $A$ and $B$ time dependent)
\begin{align}
  4\pi\left(\rho+p\right) = \frac{2\,f_{T_0}\,e^{-B}}{r} \left(A'+B'\right) \,.
  \label{016}
\end{align}
At this point we reduce our analysis to vacuum in order to find the most general  vacuum solution of the $f(T)$ field equations. We thus consider
\begin{align}
  \rho=p=0 \,,
\end{align}
from now on. Equation~(\ref{016}) immediately implies $A'+B'=0$, which we can solve for $B$
\begin{align}
  B(r) = -A(r) + \log k \,,
\end{align}
where $k$ is an integration constant. Substituting $B$ back into the field equations~(\ref{013}) (or equivalently~(\ref{014})) and~(\ref{015}) gives
\begin{align}
  & k\,r^2f_0-f_{T_0}\left( -2\,k+2\,e^A+k\,r^2\,T_0+2\,e^Ar\,A' \right) =0 \,,\\
  & 2\,k-2\,e^A+e^A\,r^2\,A'^2+e^A\,r^2\,A'' =0 \,.
\end{align}
The unique solution of these two differential equations is given by
\begin{align}
  e^{A(r)}= k \left( 1- \frac{2\,M}{r}-\frac{\Lambda_f}{3}r^2 \right) \,,
  \label{018}
\end{align}
where $M$ and $\Lambda_f$ are two constant of clear physical meaning. The value of $\Lambda_f$ depends in general from the $f(T)$ model as
\begin{align}
  \Lambda_f \equiv \frac{1}{2} \left(\frac{f_0}{f_{T_0}}-T_0\right) \,,
  \label{035}
\end{align}
which means that the cosmological constant is tuned by the $f(T)$ model under inspection. Note that this coincides with the usual cosmological constant in the TEGR limit. In other words if $f(T)\rightarrow T+2\Lambda_{GR}$ we have $\Lambda_{GR}=\Lambda_f$. Similarly it can be shown that taking the TEGR limit with $\rho=-p=8\pi\Lambda_{GR}$, the field equations lead to solution~(\ref{018}) with $\Lambda_f$ replaced by $\Lambda_{GR}$. Note that we always have to take the limit $f(T)\rightarrow T+2\Lambda_{GR}$ when we want to reduce our results to TEGR since the proper case $f(T)=T+2\Lambda_{GR}$ has been excluded from our analysis.

Whatever number the experiments will determine for $\Lambda_f$, it will not impose any constraint on the $f(T)$ model since any value of $\Lambda_f$ can be achieved fine tuning $T_0$ through a specific choice of $\gamma(r)$. This is clear looking at~(\ref{008}): since $\gamma(r)$ is an auxiliary function we can always choose it such that the value of $T_0$ is the desired one. This means that, when considering a constant torsion tensor, any $f(T)$ model can in principle admit all the solutions~(\ref{018}) parametrised by $M$ and $\Lambda_f$.

Finally we can write down the most general vacuum solution of a spherically symmetric spacetime,
\begin{align}
  ds^2=e^{A(r)}dt^2-e^{-A(r)}dr^2-r^2d\Omega^2 \,,
\end{align}
with
\begin{align}
  e^{A(r)}=  1- \frac{2\,M}{r}-\frac{\Lambda_f}{3}r^2  \,.
\label{017}
\end{align}
Note that the constant $k$ has been adsorbed with a rescaling of time, which can always be applied to static spacetimes. Solution~(\ref{017}) is nothing but the well-known SdS solution where $M$ is the Schwarzschild mass and $\Lambda_f$ the cosmological constant.
We can state that the general solutions to the spherically symmetric $f(T)$ vacuum field equations (given by tetrad~(\ref{006})) when $T=T_0=$ const, is represented by a SdS spacetime. At this point one may question whether the rotated tetrad is in fact a good choice given that it does constrain the torsion scalar. By looking at GR itself, where the Ricci scalar identically vanishes for vacuum solutions, we note that the field equations do not impose additional constraints other than those expected. In this sense we can still speak of a good tetrad.  

We also notice that equation~(\ref{015}) is identical to the isotropy condition of general relativity. This tells us that all the spherically symmetric solutions of GR will be automatically solutions of the $f(T)$ field equations~(\ref{013}),~(\ref{014}) and~(\ref{015}). However, since in Eqs.~(\ref{013}) and~(\ref{014}) appears an explicit dependence on the $f(T)$ model through $f_0$ and $f_{T_0}$, the expressions for the energy density and the pressure will differ from their general relativity counterparts.

Finally we can compare the results we found and draw a parallel with the ones obtained within the more common $f(R)$ theories of gravity. In analogy with what we exposed above, solutions to the Einstein field equations have been extended to $f(R)$ theories when the Ricci scalar is constrained to be constant~\cite{Benachour:2012hh}. Restricting to spherical symmetry, it has been shown that the Schwarzschild-de Sitter solution is a solution of the $f(R)$ field equations whenever $R$ is constant~\cite{Multamaki:2006zb,Multamaki:2006ym,Sebastiani:2010kv}. Interestingly enough, in these solutions the cosmological constant is related to the Ricci scalar in analogy to how $\Lambda_f$ is connected to the torsion scalar through~(\ref{035}). This shows that $T$ and $R$ play similar roles in the respective theories underlining the fundamental difference between GR and TEGR, namely the description of spacetime by curvature or torsion.

Furthermore it is well known that within metric $f(R)$ gravity Birkhoff's theorem holds only if some constraint are imposed on the Ricci scalar $R$, such as requiring it to be constant~\cite{Capozziello:2007ms,Capozziello:2007id,Faraoni:2010rt}. On the other hand,  Palatini $f(R)$ gravity in vacuum reduces to GR plus a cosmological constant and $R$ becomes automatically constant. Birkhoff's theorem is thus valid and all the GR solutions are also solutions of the modified field equations~\cite{Kainulainen:2007bt,Kainulainen:2006wz,Barausse:2007pn,Barausse:2007ys}. We can compare this with our analysis on non-diagonal (rotated) tetrads in $f(T)$ gravity. When the torsion scalar is constrained to be constant we managed to prove Birkhoff's theorem and to construct SdS solutions in strict analogy with both metric and Palatini $f(R)$ theories. In the particular case of Sec.~\ref{appA} we even prove the theorem without imposing constraints on $T$, which seem to go beyond the results of $f(R)$ gravity. However in this case the Schwarzschild solution is not a solution of the $f(T)$ field equations (see~\cite{Boehmer:2011gw}) and the chances to find physical applications are small.

In conclusion, all the results we found within $f(T)$ gravity have similar counterparts in $f(R)$ gravity. This suggests that all the achievements for spherically symmetric spacetime already obtained in $f(R)$ theories can be similarly transposed to $f(T)$ theories with much work left for future studies.

\section{Cosmology in spherical coordinates}
\label{Sec:Cosmo}

Cosmological applications of $f(T)$ gravity have a short but productive history beginning with~\cite{Ferraro:2006jd,Ferraro:2008ey}. Models have been built to explain both early and late time accelerated expansion~\cite{Myrzakulov:2010vz,Myrzakulov:2010tc, Yerzhanov:2010vu, Yang:2010hw, Karami:2010bu, Farajollahi:2011af, Wu:2010av, Bamba:2010wb, Zhang:2011qp} and several issues have been recently analysed \cite{Liu:2012kk,Chen:2010va, Yang:2010ji, Dent:2011zz, Cai:2011tc, Chattopadhyay:2011fp, Wei:2011jw, Wei:2011mq, Capozziello:2011hj, Bamba:2011pz, Karami:2011nj, Wei:2011aa, Atazadeh:2011aa, Karami:2012fu, Karami:2012if, Setare:2012vs, Zheng:2010am, Li:2011wu, Wu:2010mn, Bengochea:2010sg, Wu:2011xa,Fu:2012ix}.

In $f(T)$ cosmology, considering the Friedmann-Lemaitre-Robertson-Walker (FLRW) metric in Cartesian coordinates the diagonal tetrad seems to be a good gauge choice since it leads to field equations not containing constraints on the $f(T)$ function or on the torsion scalar $T$. However when choosing the diagonal tetrad in spherical coordinates, we get the unwanted condition $f_{TT}=0$, which is satisfied by TEGR only. We will demonstrate using rotated tetrads that it becomes possible to built a well-defined tetrad for FLRW cosmology in spherical coordinates. This allows us to generalise $f(T)$ gravity to the non spatially flat FLRW cosmologies, which cannot be done easily with Cartesian coordinates and has only been considered using hyperspherical coordinates~\cite{Ferraro:2011us,Ferraro:2011zb}.

Let us start with the spatially flat FLRW line element in Cartesian coordinates
\begin{align}
  ds^2=dt^2-a(t)^2\left(dx^2+dy^2+dz^2\right)\,,
\label{101}
\end{align}
where $a(t)$ is the usual scale factor. It is well known and easy to verify that the diagonal tetrad
\begin{align}
  e^a_\mu=\mbox{diag}\left(1,a(t),a(t),a(t)\right)\,,
\label{102}
\end{align}
results in~(\ref{101}) and leads to the following field equations
\begin{align}
  T &= -6\,H^2\,, 
  \label{108}\\ 
  4 \pi \rho &= 3 H^2 f_T+\frac{1}{4} f\,,
  \label{111}\\
  4 \pi \left(p+ \rho\right) &= \dot H \left(12 H^2 f_{TT}- f_T\right) \,, 
  \label{112}
\end{align}
where $H(t)\equiv \dot a / a$ is the Hubble parameter. The diagonal tetrad~(\ref{102}) seems to represent a good choice among all the possible tetrads giving metric~(\ref{101}) since it yields to a modification of the analogous GR field equations which do not involve any constraint on either the function $f(T)$ or the torsion scalar $T$. As we are going to see, this is not the case when using spherical coordinates.

Consider now the FLRW line element in spherical coordinates
\begin{align}
  ds^2=dt^2-a(t)^2\left(\frac{dr^2}{1-k\,r^2}+r^2d\theta^2+r^2\sin^2\theta\, d\phi^2\right)\,,
  \label{103}
\end{align}
where $k=0,+1,-1$ correspond to the flat, closed or open FLRW spacetime. We therefore expect that our tetrad rotations may allow us to identify `good' tetrads in the context of cosmology. Again, the simplest tetrad returning metric~(\ref{103}) is the diagonal one
\begin{align}
  e_\mu^a=\mbox{diag}\left(1,\frac{a(t)}{\sqrt{1-k\,r^2}},a(t)\,r,a(t)\,r\,\sin\theta\right) \,.
  \label{104}
\end{align}
However the (off-diagonal) $f(T)$ field equations leads in this case to the condition
\begin{align}
  f_{TT}=0 \,,
\end{align}
which is satisfied if and only if $f(T)=c_1 T+c_2$, i.e. TEGR with a cosmological constant. This happens independently of the value of the spatial curvature parameter $k$. It is then clear that tetrad~(\ref{104}) does not represent a useful tetrad in $f(T)$ cosmology since, although it correctly gives the FLRW metric~(\ref{103}), it leads to an unwanted constraint over all the possible $f(T)$ models which set the theory to be TEGR.

We now see how we can define a useful tetrad making a local rotation. Let us focus on the spatially flat ($k=0$) cosmology for the moment and look at tetrad~(\ref{103}). We know that any local Lorentz transformation ${\Lambda^a}_b$ can be applied to tetrad~(\ref{103}) without altering metric~(\ref{101}). That is the relation
\begin{align}
  \eta_{ab}\,{e_\mu}^a\,{e_\nu}^b=g_{\mu\nu} \,,
\end{align}
is unchanged by the transformation
\begin{align}
  {e_\mu}^a\rightarrow{\Lambda^a}_b\,{e_\mu}^b \,.
\end{align}
As before, consider a general 3-dimensional rotation $\mathcal{R}$ in the tangent space parametrised by the three Euler angles $\alpha$, $\beta$, $\gamma$
 \begin{align}
  {\Lambda^a}_b=
  \begin{pmatrix}
    1 & 0 \\
    0 & \mathcal{R}(\alpha,\beta,\gamma) \\
  \end{pmatrix}\,.
  \label{106}
\end{align}
We reduce this transformation considering the following values for the three Euler angles
\begin{align}
  \alpha=\theta-\frac{\pi}{2}\,, \qquad \beta=\phi\,, \qquad \gamma=\gamma(t,r) \,,
\label{125}
\end{align}
where $\gamma$ is taken to be a general function of both $t$ and $r$. In this manner, applying the rotation matrix~(\ref{106}) to the diagonal tetrad~(\ref{103}) with $k=0$ gives the following rotated tetrad explicitly
\begin{multline}
  {e_\mu}^a=
  \left(
  \begin{array}{cc}
    1 & 0 \\
    0 & a \sin \theta \cos \phi  \\
    0 & -a\, r (\cos \theta \cos \phi \sin \gamma+\sin \phi \cos \gamma) \\ 
    0 & a\, r \sin \theta (\sin \phi \sin \gamma-\cos \theta \cos \phi \cos \gamma)
  \end{array}\right. \\
  \left.
  \begin{array}{cc}
    0 & 0 \\
    a \sin \theta \sin \phi  & a \cos \theta  \\
    a\, r (\cos \phi \cos \gamma-\cos \theta \sin \phi \sin \gamma) & a\, r
    \sin \theta \sin \gamma \\
    -a\, r \sin \theta (\cos \theta \sin \phi \cos \gamma+\cos
    \phi \sin \gamma) & a\, r \sin ^2\theta \cos \gamma
  \end{array}
  \right) \,.
  \label{107}
\end{multline}
With this tetrad the torsion scalar reads
\begin{align}
  T(t,r)=\frac{4\, r\, \gamma ' \cos \gamma+4 \sin\gamma-6\, r^2\, \dot a^2+4}{r^2 a^2} \,.
\label{126}
\end{align}
From the off-diagonal $f(T)$ field equations we obtain the constraints $\dot T\,f_{TT}=T'\,f_{TT}=0$, which force the $f(T)$ model to be TEGR again unless we have a constant $T$. The mathematical reason for this to happen is that the torsion scalar depends on both $t$ and $r$. We may try to eliminate this dependency on one of the two coordinates and see if a non-constant $T$ is allowed. We have then to choose $\gamma(t,r)$ such that $T$ will depend only on $t$ or $r$.

\subsection{$T$ depending on $r$ only}

First we analyse the case $T(t,r)=T(r)$. In order to eliminate the time dependence in $T$ we can make the following choice for $\gamma$
\begin{align}
  \sin\gamma(t,r)\equiv -1+\frac{1}{2}r^2 \dot a^2+ \psi(r)\, a^2 \,,
  \label{115}
\end{align}
where $\psi$ is an arbitrary function of $r$. The torsion scalar now becomes
\begin{align}
  T(r)=\frac{4 \left(r\, \psi '+\psi\right)}{r^2} \,,
\end{align}
which is clearly independent of $t$. Note that we must require
\begin{align}
  \left|-1+\frac{1}{2}r^2 \dot a^2+ \psi(r)\, a^2\right| \leq 1 \,,
\end{align}
because of definition~(\ref{115}). This cannot be true everywhere since at any fixed time the second term is larger than 1 if $r$ is sufficiently big. However, requiring $\psi$ and $\dot a$ to be sufficiently small, we can satisfy this condition everywhere but for $r\rightarrow\infty$. Note that this poses physical constraints on the Hubble parameter. In any case, from the $f(T)$ field equations we get the condition
\begin{align}
  T'f_{TT}=0 \,,
\end{align}
which either reduce the analysis to TEGR or forces $T$ to be constant. Excluding the first case we can see what happens when $T=T_0=$ constant. The remaining two independent field equations becomes
\begin{align}
  16\pi\rho &= f_0 -f_{T_0} \left(T_0-6H^2\right) \label{116} \,,\\
  4\pi\,(\rho+p) &=- \dot H f_{T_0} \,, \label{117}
\end{align}
where $f_0=f(T_0)$ and $f_{T_0}=f'(T_0)$ are constants.  We can find conditions when the universe is accelerating in this case analysing the acceleration condition $\ddot a > 0$. This is satisfied whenever
\begin{align}
  \frac{\rho+3\,p}{f_{T_0}} < -\frac{3}{4\pi}\left(\frac{f_0}{f_{T_0}}-T_0\right) \equiv-\frac{3}{2\pi} \Lambda_f \,,
\label{118}
\end{align}
or, in terms of the equation of state parameter $w$,
\begin{align}
\frac{w}{f_{T_0}}< -\frac{\Lambda_f}{2\pi\rho}-\frac{1}{3f_{T_0}} \,.
\end{align}
Notice that when we reduce to TEGR, for which $f_{T_0}=1$ and $\Lambda_f=0$, we correctly get $w<-1/3$. Unfortunately, eliminating the time dependence we cannot eliminate of the $T=$ const condition.

\subsection{$T$ dependent on $t$ only}

Let now see what happens if we eliminate the radial dependence. Choosing
\begin{align}
  \sin\gamma(t,r)\equiv -1+\frac{\xi(t)}{r}+r^2\,\chi(t)\,,
  \label{109}
\end{align}
with $\xi$ and $\chi$ general functions of $t$. We find
\begin{align}
  T(t,r)=T(t)=12\,\frac{\chi}{a^2}-6 H^2 \,,
  \label{129}
\end{align}
which is indeed independent of $r$. The function $\chi(t)$ represents a modification with respect to the torsion scalar found in Cartesian coordinates~(\ref{108}). Note that definition~(\ref{109}) implies
\begin{align}
  \left|-1+\frac{\xi(t)}{r}+r^2\,\chi(t)\right|\leq 1 \,,
  \label{110}
\end{align}
which is identically satisfied if and only if $\xi=\chi=0$. However, given $\xi$ and $\chi$ positive and sufficiently small, we can satisfy~(\ref{110}) everywhere but in the neighbourhoods of $r=0$ and $r\rightarrow\infty$.

The independent $f(T)$ field equations we obtain in this case are
\begin{align}
  16 \pi \rho &=-\left(T-6 H^2\right) f_T+f \label{113} \,,\\
  4 \pi\,( p+\rho)&=- \dot T H f_{TT}-\dot H f_T \label{114} \,,\\
0&= \dot T f_{TT} \left(r^3 \chi+\xi\right) \,. \label{130}
\end{align}
The first two of these equations coincide with~(\ref{111}) and~(\ref{112}) when the torsion scalar reduces to $T=-6H^2$, which happens for $\chi(t)=0$. The last equation generalises the off-diagonal constraint and it is satisfied if $f_{TT}=0$, $\dot T=0$ or $\xi=\chi=0$. The first case leads us again back to TEGR and is thus not of interest. The second case gives $T=T_0=$ const and reduces the remaining equations to
\begin{align}
  16 \pi \rho &=-\left(T_0-6 H^2\right) f_{T_0}+f_0 \,,\\
  4 \pi\,( p+\rho)&=-\dot H f_{T_0} \,,
\end{align}
where $f_0=f(T_0)$ and $f_{T_0}=f'(T_0)$ are also constants. These equations are equal to~(\ref{116}) and~(\ref{117}), meaning that they describe the same physical system. In particular we would recover the acceleration condition~(\ref{118}).

The last case requires $\xi=\chi=0$ and thus $\sin\gamma=-1$ which implies $\gamma=-\pi/2+2n\pi$ with $n$ an integer number. Tetrad~(\ref{107}) becomes equivalent to the well-known rotated tetrad used in~\cite{Boehmer:2011gw}. More interestingly in this case we do not gain any constraint on the $f(T)$ model or on the torsion scalar $T$ which takes the form~(\ref{108}). Equations~(\ref{113}) and~(\ref{114}) reduce to equations~(\ref{111}) and~(\ref{112}), respectively, meaning that the theory becomes completely equivalent to its Cartesian counterpart. We thus managed to find a tetrad which allows a general (no constraints on $f(T)$ or $T$) description of the universe which is physically equivalent to the Cartesian coordinates analysis. This tetrad is given performing the same rotation considered in~\cite{Boehmer:2011gw} and reads
\begin{align}
  {e^{a}}_\mu =
  \begin{pmatrix}
    1 & 0 & 0 & 0 \\
    0 & a(t) \sin\theta\cos\phi  & a(t)\, r\cos \theta\cos\phi & - a(t)\, r\sin \theta\sin \phi \\
    0 & a(t) \sin\theta\sin\phi & a(t)\, r\cos \theta\sin\phi& a(t)\, r\sin\theta \cos\phi \\
    0 & a(t) \cos \theta  & - a(t)\, r \sin\theta & 0
  \end{pmatrix}\,.
  \label{119}
\end{align}

\subsection{Spatially curved FLRW $f(T)$ cosmology}

Finally, we analyse the non spatially flat ($k\neq 0$) FLRW universe described by metric~(\ref{103}). We can apply the rotation matrix~(\ref{106}) to the diagonal tetrad~(\ref{104}) with the Euler angles values~(\ref{125}).
The torsion scalar now becomes
\begin{align}
  T=\frac{4}{r^2 a^2} \left( r \sqrt{1-k r^2} \gamma ' \cos\gamma+4 \sqrt{1-k r^2} \sin\gamma-6 r^2 \dot a^2-2 k r^2+4 \right) \,,
  \label{127}
\end{align}
generalising~(\ref{126}) which is recovered setting $k=0$. Again we want to reduce $T$ to become function of only $t$ or $r$. Since we expect the $T'=0$ constraint as for the $k=0$ universe, we will only analyse the case where $T=T(t)$. In order to reduce~(\ref{127}) to be $r$ independent we choose a particular expression for $\gamma(t,r)$, namely
\begin{multline}
  -\sin\gamma = \frac{1}{4 r}\biggl(\frac{3 \arcsin(\sqrt{k} r) (k-2 \chi (t))}{k^{3/2}} \\
  + \frac{r \sqrt{1-k r^2} (k+6 \chi (t))}{k} - 4 \xi (t) \biggr) \,,
  \label{128}
\end{multline}
which in the limit $k\rightarrow 0$ becomes~(\ref{109}). Here $\chi$ and $\xi$ are two general functions of $t$. Of course in order for~(\ref{128}) to be consistent we must require that the right-hand side of~(\ref{128}) is between -1 and 1. This in general poses constraints on the functions $\chi$ and $\xi$. We will see in what follows that this issue will play a main role if the universe is open ($k=-1$).
The torsion scalar~(\ref{127}) is given by
\begin{align}
  T=12\frac{\chi}{a^2}-6 H^2 \,,
\end{align}
which depends only on $t$ and coincides with~(\ref{129}).

The independent field equations are
\begin{align}
  16 \pi\rho =f+f_T\left(6\frac{k }{a^2}+6 H^2 -T\right) \,, \\
  4 \pi( p+\rho) = f_T\left(\frac{k}{a^2}- \dot H\right)-\dot T H f_{TT} \,,\\
  \frac{\dot{T} f_{TT}}{k^{3/2}} \left[4 k^{3/2} \xi+3\, (k-2 \chi)\left(\sqrt{k} r \sqrt{1-k r^2}-\arcsin\left(\sqrt{k} r\right)\right) \right]=0 \,, \label{131}
\end{align}
which reduces to~(\ref{113})--(\ref{130}) in the limit $k\rightarrow 0$. There are three ways to satisfy~(\ref{131}): $f_{TT}=0$, $\dot T$=0 or both
\begin{align}
  \xi=0 \quad\mbox{and}\quad \chi=k/2 \,.
  \label{132}
\end{align}
The first possibility takes us back to TEGR and thus is of no interest. In the second case we have $T=$ const and the remaining field equations become
\begin{align}
  16 \pi\rho &= f_0+f_{T_0}\left(6\frac{k }{a^2}+6 H^2 -T_0\right) \,, 
  \label{135} \\
  4 \pi( p+\rho) &= f_{T_0}\left(\frac{k}{a^2}- \dot H\right) \,, 
  \label{136}
\end{align}
which are the $k\neq 0$ counterparts of~(\ref{116}) and~(\ref{117}), respectively.

Finally, in the last and more interesting case, the torsion scalar becomes
\begin{align}
  T=6\,\frac{k}{a^2}-6\,H^2 \,,
\end{align}
and the field equations reduce to
\begin{align}
  4\pi\rho &= \frac{f}{4} +3\, H^2 f_T \label{121} \,,\\
  4\pi\,(\rho+p) &= -f_T \left(\dot H-\frac{k}{a^2}\right)+ 12\,H^2f_{TT} \left(\dot H+\frac{k}{a^2}\right) \,, \label{122}
\end{align}
which clearly generalises equations~(\ref{111}) and~(\ref{112}). Note that equations~(\ref{121}) and~(\ref{122}) correctly reduce to the usual Friedmann equations when $f(T)=T$. It seems thus that the choice~(\ref{132}) allows us to generalise FLRW cosmology to its spatially curved cases without giving rise to any constraint regarding $T$ or the $f(T)$ function. Unfortunately this does not work for the open FLRW universe as we are going to explain.

For the values~(\ref{132}) the function $\gamma(t,r)$ as given by~(\ref{128}) reduces to
\begin{align}
  \gamma=-\arcsin\sqrt{1-kr^2} \,.
  \label{133}
\end{align}
In order for this to be mathematically consistent we must require
\begin{align}
  \left|\sqrt{1-kr^2}\right| \leq 1 \,.
  \label{134}
\end{align}
Since $r^2\geq 0$, when $k=-1$ this condition is never satisfied, meaning that the open FLRW universe cannot be described by a tetrad containing~(\ref{133}). The mathematical reason of this issue is that~(\ref{127}) has no (real) solutions reducing to~(\ref{133}) when $k=-1$. It seems tempting to replace the trigonometric function by a hyperbolic function, however, this is not allowed since the resulting transformation would no longer be a local Lorentz transformation. On the other hand, in the closed FLRW universe condition~(\ref{134}) is always satisfied because $r\leq 1$ when $k=1$.

The entire issue can be easily understood switching to {\it hyper-spherical coordinates} which transforms the radial coordinate $r$ to $\psi$ as
\begin{align}
  r\mapsto \frac{\sin(\sqrt{k}\psi)}{\sqrt{k}} =
  \left\{
  \begin{array}{ll}
    \sin\psi & \mbox{if } k=1 \\
    \psi & \mbox{if } k\rightarrow 0 \\
    \sinh\psi & \mbox{if } k=-1
  \end{array}
  \right. \,.
\end{align}
The FLRW line element now reads
\begin{align}
  ds^2=dt^2-a^2 \Bigl(d\psi^2+\frac{\sin^2(\psi\sqrt{k})}{k}d\Omega^2 \Bigr) \,,
\end{align}
while condition~(\ref{134}) becomes
\begin{align}
  \left|\cos\left(\psi\sqrt{k}\right)\right|\leq 1 \,,
\end{align}
which is always satisfied when $k=1$ since clearly $\cos \psi \leq 1$. In the $k=-1$ case instead we find $\cosh \psi\leq 1$ which is never satisfied unless $\psi=0$.

In conclusion we firstly stress that the diagonal tetrad~(\ref{104}) cannot be used in $f(T)$ cosmology since it forces the $f(T)$ model to be TEGR. We can state that in the closed FLRW universe the rotated tetrad containing~(\ref{133}), represents a good choice among all the possible tetrads giving back metric~(\ref{104}). It does not imply any constraint on either the functional form of $f(T)$ or on the torsion scalar $T$ itself. It does lead to a simple generalisation of Eqs.~(\ref{111}) and~(\ref{112}) which reduce to the correct Friedmann equations in the limit $f(T)\rightarrow T$. The `good' tetrad for $k=1$ cosmology thus reads
\begin{multline}
  {e_\mu}^a =
  \left(
  \begin{array}{cc}
    1 & 0  \\
    0 & a \cos\phi  \sin\theta /\sqrt{1-r^2} \\
    0 & r\, a \left(\sqrt{1-r^2} \cos \theta  \cos \phi -r \sin\phi \right)  \\
    0 & r\, a \sin\theta  \left(-r \cos\theta  \cos \phi -\sqrt{1- r^2} \sin\phi\right)
  \end{array}\right.\\
  \left.
  \begin{array}{cc}
    0 & 0 \\
    a \sin \theta  \sin \phi /\sqrt{1- r^2} & a \cos \theta /\sqrt{1- r^2} \\
    r\, a \left(r \cos\phi +\sqrt{1- r^2} \cos\theta  \sin\phi
    \right) & -r \sqrt{1- r^2}\, a \sin\theta \\
    r\, a \sin \theta  \left(\sqrt{1- r^2} \cos \phi -
    r \cos \theta  \sin \phi \right) & r^2\, a \sin ^2\theta
  \end{array}
  \right) \,,
  \label{123}
\end{multline}
or in hyperspherical coordinates
\begin{multline}
  {e_\mu}^a =
  \left(
  \begin{array}{cc}
    1 & 0 \\
    0 & a \cos\phi \sin\theta \\
    0 & a \sin\psi \left(\cos\psi \cos\theta \cos\phi\pm\sin\psi \sin\phi \right) \\
    0 & -a \sin\psi \sin\theta \left(\pm\cos\theta \sin\psi \cos\phi+\cos\psi \sin\phi\right)
  \end{array}
  \right.\\
  \left.
  \begin{array}{cc}
    0 & 0 \\
    a \sin\theta \sin\phi & a \cos\theta \\
    a \sin\psi \left(\pm\sin\psi \cos\phi+\cos\psi \cos\theta \sin\phi\right) & -a \cos\psi \sin\psi \sin\theta \\
    a \sin\psi \sin\theta \left(\cos\psi \cos\phi \pm\cos\theta \sin\psi \sin\phi \right) & \pm a \sin\psi \sin\psi \sin ^2\theta 
  \end{array}
  \right) \,,
\end{multline}
where the $\pm$ signs come from the square root terms.

On the other hand, when the FLRW universe is open it seems impossible, using the methods explored in this paper, to find a tetrad having all the properties above. The best we can do is to leave the functions $\xi(t)$ and $\chi(t)$ in~(\ref{128}) undetermined and deal with field equations~(\ref{135}) and~(\ref{136}). However in this case the torsion scalar is constrained to be constant and we can only achieve a theoretically limited descriptions of the universe.

Condition~(\ref{134}) seems to suggest that a complex rotation might work. This idea has been successfully used in~\cite{Ferraro:2011us,Ferraro:2011zb} where the authors were able to construct a good tetrad for the hyperbolic FLRW universe by introducing complex tetrads. Another possibility which might avoid the use of complex tetrads could be the use local Lorentz boosts, instead of rotations, see~\cite{Ferraro:2011ks} where boosts have been used to construct very useful tetrads in the context of spherical symmetry. 

\section{Conclusions}
\label{Sec:Concl}

This paper emphasised the importance of choosing appropriate tetrads when analysing the field equations of $f(T)$ gravity. This issue arises since the theory is not invariant under local Lorentz transformations and thus different tetrads will yield different field equations which in turn might have different solutions. Some of these solutions do not have a valid GR counterpart, while others tend to their GR counterparts in the appropriate limit. Therefore, special attention has to be given to the choice of tetrad.

We thus introduce the notion of a good tetrad. By a good tetrad we mean a tetrad which gives rise to field equations which do not constrain the functional form of $f(T)$. In such cases one can always consider the limit $f(T) \rightarrow T$ where the correct general relativistic limit is recovered. Otherwise we will talk of a bad tetrad. It is well known that the diagonal tetrad is bad when working in spherical symmetry. Our rotated tetrads and also the boosted tetrad of~\cite{Ferraro:2011ks} are good tetrads according to our notion. 

We were able to construct good tetrads in spherical symmetry by simply rotating the diagonal tetrad. The power of this simple method to generate potentially good tetrads was shown by proving Birkhoff's theorem in $f(T)$ gravity and thereby showing that the SdS class of solutions is the unique vacuum solution of the vacuum field equations. We suggest to revisit static and spherically symmetric perfect fluid solutions along the lines of~\cite{Boehmer:2011gw} as it should be straightforward to generate a large number of new exact solutions of the field equations with perfect fluid source.

We extended our programme of rotating tetrads to the study of homogeneous and isotropic spacetimes, i.e.~cosmology. We were able to construct good tetrads not only for the spatially flat metric but also for the closed universe. In both cases there are no constraints on the functional form of $f(T)$ and it is possible to study a very general class of $f(T)$ cosmologies.

In principle, our approach of rotating tetrads in the tangent space might be applicable to situations with other symmetries. Recall that the Lorentz transformations form a six-dimensional group, three boosts and three rotations. Therefore, one can introduce six auxiliary fields into the field equations, simply be performing a general Lorentz transformation on the tetrad. Then the aim is to eliminate those field equations which yield constraints on $f(T)$. We achieved this by considering rotations in spherical symmetry. However, we are confident to say that there are enough degrees of freedom in these transformations to always find a good tetrad according to our definition. Of course, there may be situations where one wishes to enforce other properties by fixing the Lorentz transformations accordingly.

\end{document}